\documentclass[12pt]{iopart}
\usepackage[utf8]{inputenc} 
\usepackage{graphicx}
\usepackage{color}

\begin{document}

%\graphicspath{{Submission/}} % remove for submission

\title[In content and emission wavelength of In$_x$Ga$_{1-x}$N/GaN nanowires]%
{Correlation between In content and emission wavelength of In$_{x}$Ga$_{1-x}$N/GaN nanowire heterostructures}

\author{M Wölz, J Lähnemann, O Brandt, V M Kaganer, M Ramsteiner, C Pfüller,
C Hauswald, C N Huang, L Geelhaar and H Riechert}

\address{Paul-Drude-Institut für Festkörperelektronik,
Hausvogteiplatz 5--7, 10117 Berlin, Germany}

\ead{woelz@pdi-berlin.de}

\begin{abstract}
GaN nanowire ensembles with axial In$_x$Ga$_{1-x}$N multi-quantum wells (MQWs) were grown by molecular beam epitaxy. In a series of samples, we varied the In content in the MQWs from almost zero to about 20\,\%. Within the nanowire ensemble, the MQWs fluctuate strongly in composition and size. Statistical information about the composition was obtained from x-ray diffraction and Raman spectroscopy. Photoluminescence at room temperature was obtained in the range from 2.2\,eV to 2.5\,eV depending on In content. Contrary to planar MQWs, the intensity increases with increasing In content. We compare the observed emission energies with transition energies obtained from a one-dimensional model, and conclude that several mechanisms for carrier localization affect the luminescence of these three-dimensional structures.
\end{abstract}


\submitto{\NT}
%\submitto{\NT \today} % remove for submission

%\maketitle

%%%%%%%%%%%%%%%
\section{Introduction}
%%%%%%%%%%%%%%%

GaN nanowires (NWs) are pursued as an alternative to planar GaN for the fabrication of light emitting diodes (LEDs) \cite{Kikuchi2006, Guo2010, Bavencove2010, Armitage2010, Lin2010, Nguyen2011a}. The advantage of GaN NWs is that they can be synthesized on cost-effective Si substrates in excellent crystal quality \cite{Geelhaar2011}.
NWs also have the benefit over planar layers that strain from lattice-mismatched heterostructures can elastically relax at the free sidewalls and defects from plastic relaxation can be avoided \cite{Bjoerk2002}.
Several groups have fabricated GaN NWs, with repeated axial In$_x$Ga$_{1-x}$N segments to form multi-quantum-wells (MQWs), on Si substrates by plasma-assisted molecular beam expitaxy (PA-MBE).
Luminescence with peak emission wavelengths from 440\,nm to 640\,nm has been reported \cite{Kikuchi2006, Lin2010, Nguyen2011a, Chang2010}.
For the application of such MQWs in devices like LEDs it is mandatory to control the emission wavelength.
Still, at present, it is even unclear how the emission of ensembles of axial In$_x$Ga$_{1-x}$N/GaN NW heterostructures varies with the In content.
The luminescence wavelength is known to increase at lower growth temperatures of the active NW region, because more In is incorporated \cite{Nguyen2011a}.
However, there are only few reports on the determination of the In content in NW MQWs which do not rely on the luminescence wavelength and therefore assume its dependence on the In content \emph{a priori} \cite{Chang2010,Knelangen2011}. Moreover, in these reports the In content has been quantified only for single NWs, and the effect of a systematic variation of the In content has not been shown.
There are two major experimental difficulties. First, for NWs it is significantly more difficult to determine the MQW structural parameters than for planar layers. Second, these parameters fluctuate significantly on a given sample.
Analytical tools are required which probe large ensembles of NWs and deliver statistical information on the MQW structure. Recently, we have shown that the composition and thickness fluctuations of In$_x$Ga$_{1-x}$N/GaN MQWs in self-induced NWs, which form superlattices if they are equally spaced, can be determined by laboratory x-ray diffraction (XRD) \cite{Kaganer2011a, Woelz2011b, Woelz2012}.
Here, we make use of this method and analyze a series of In$_x$Ga$_{1-x}$N/GaN NW samples for which the growth temperature of the active region and consequently the average In content was varied. The results for the In concentration are corroborated by resonant Raman spectroscopy. We then compare the observed photoluminescence (PL) with the expected transition energies calculated by a one-dimensional model.

%%%%%%%%%%%%%%%
\section{Sample growth}
%%%%%%%%%%%%%%%

\begin{figure}
\hspace{1in}\includegraphics{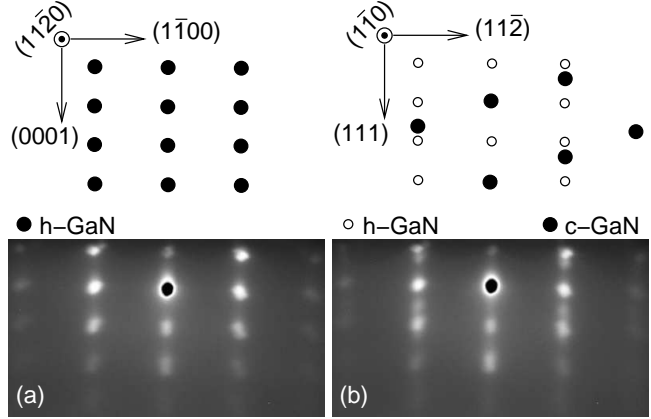}
\caption{\label{fig1}Evolution of RHEED pattern during growth of GaN NWs (a) in the base and (b) in the active region at reduced growth temperature, before In$_x$Ga$_{1-x}$N is grown. The appearance of the cubic GaN diffraction pattern is depicted above for clarity (one of the twin patterns drawn).}
\end{figure}

GaN NWs were grown on Si(111) by PA-MBE at 770\,$^{\circ}$C with a N/Ga flux ratio of 4. After one hour of growth, the metal flux was interrupted and the substrate temperature was lowered for the growth of the active region, which is formed by six identical pairs of GaN barriers and In$_x$Ga$_{1-x}$N QWs. The achieved nominal thicknesses were 14\,nm for the barriers and 5\,nm for the QWs.
During the initial phase of the growth of the active region, evidence for the formation of cubic GaN (c-GaN) was obtained \emph{in situ} by reflection high-energy electron diffraction (RHEED). Figure~\ref{fig1} (a) shows the RHEED pattern during the growth of the GaN NW base, when only the reflections of hexagonal GaN (h-GaN) were observed. The appearance of additional c-GaN reflections after lowering the growth temperature is depicted in figure \ref{fig1} (b). The inclusions of c-GaN affect the optical properties of the NWs, as we will see later.

A series of five samples was produced with the growth temperature of the active region, $T_a$, ranging from 583\,$^{\circ}$C to 637\,$^{\circ}$C while all fluxes and shutter opening sequences were kept unchanged. At these temperatures, the desorption of In cannot be neglected. Therefore, the resulting composition $x$ depends on $T_a$ \cite{Averbeck1999}.
Scanning electron microscopy has revealed that the resulting NW diameters fluctuate, with typical values between 30\,nm and 100\,nm (images not shown).

%%%%%%%%%%%%%%%
\section{Structural analysis}
%%%%%%%%%%%%%%%
%%%%%%%%%%%%%%%
\subsection{X-ray diffraction}
%%%%%%%%%%%%%%%

\begin{figure}
\hspace{1in}\includegraphics{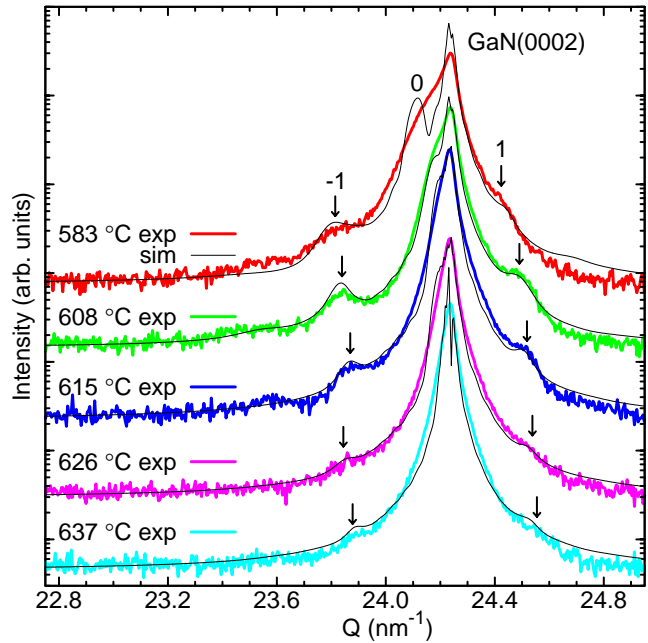}
\caption{\label{fig2}Experimental and simulated $\omega$-$2\theta$ scans across the GaN(0002) reflection of In$_x$Ga$_{1-x}$N/GaN NW samples grown at different temperatures $T_a$. The periodic In$_x$Ga$_{1-x}$N QWs give rise to the first order superlattice peaks as indicated by the arrows.}
\end{figure}

In order to determine the In concentration, we performed x-ray measurements
with CuK$\alpha_{1}$ radiation using a Ge(220) hybrid monochromator and a Ge(220) analyzer crystal. Symmetric
Bragg scans across the GaN(0002) reflection are
shown in figure \ref{fig2}.
The zeroth order satellite peak in the left shoulder of the GaN(0002) reflection indicates the increased average lattice spacing $c_\mathrm{avg}$ of the MQW region due to In incorporation. This peak is surrounded by regularly spaced satellites on both sides, the $\pm 1^\textrm{st}$ orders of which are marked by
arrows in figure \ref{fig2}. These satellites arise from thickness interference at
the periodic QW stack, and their positions can be used to determine the
superlattice period $d_{\mathrm{SL}}$ and $c_{\mathrm{avg}}$ via Bragg's law \cite{Woelz2011b}. We obtain $d_{\mathrm{SL}}=19\pm1$\,nm for these samples.
For axial superlattices in NWs, lateral relaxation occurs. The XRD peak positions are determined by the average c-plane lattice parameters in the QW and barrier. For the interpretation of the XRD profiles, the following simplification may therefore be made: If the total height of the superlattice stack is larger than the NW diameter, the whole stack assumes a weighted average in-plane lattice constant \cite{Kaganer2012}. In this case the Poisson effect vanishes for $c_{\mathrm{avg}}$ \cite{Woelz2012, Kaganer2012}. The average superlattice composition is hence given by Vegard's law: $x_\mathrm{avg} = (c_\mathrm{avg}-c_\mathrm{GaN})/(c_\mathrm{InN}-c_\mathrm{GaN})$.

%%%%%%%%%%%%%%%
\subsection{TEM}
%%%%%%%%%%%%%%%

\begin{figure}
\hspace{1in}\includegraphics{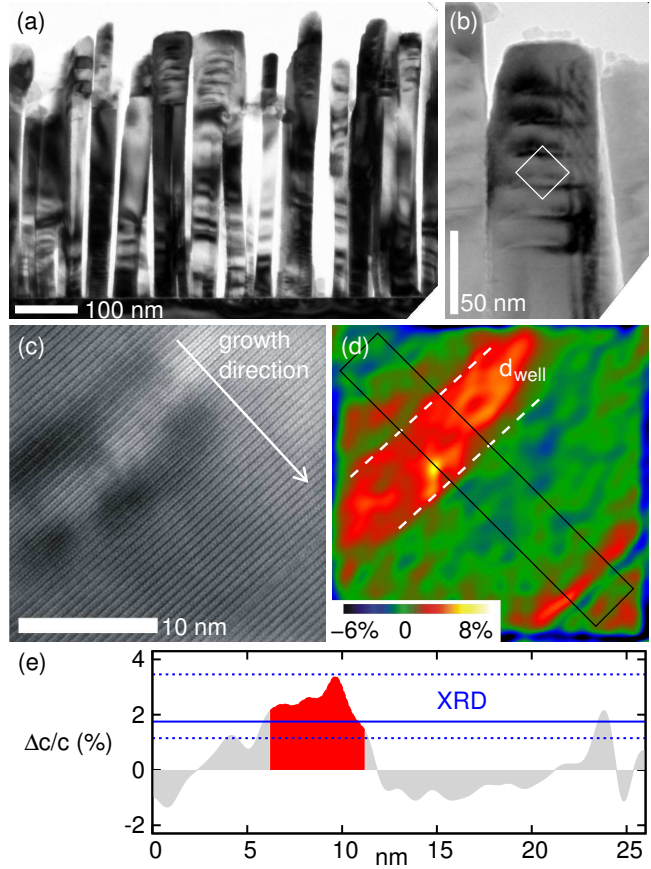}
\caption{\label{fig3}TEM analysis of axial In$_x$Ga$_{1-x}$N QW in GaN NWs. (a) NW ensemble in bright field contrast. (b) NW tip with embedded MQW. (c) HRTEM image of one QW, from the area marked by the box in (b). (d) $\Delta c/c$ map obtained from GPA, with the estimated QW thickness marked by dashed lines. (e) $\Delta c/c$ strain profile along the NW axis, obtained within the black box in (d). The approximate location of the QW is marked red, and the QW strain estimate from XRD is given by the blue line.}
\end{figure}

In the experiments shown in figure \ref{fig2}, the contrast of the XRD satellite peaks is not sufficient to determine the QW thickness $d_\mathrm{well}$. From a calibration of the axial In$_x$Ga$_{1-x}$N growth rate \cite{Woelz2011b}, we expect $d_\mathrm{well} = 5$\,nm. 
To cross-check the structure of the NW samples, TEM was performed on the sample grown at $T_a = 608$\,$^{\circ}$C. Figure \ref{fig3} (a) shows a bright-field image of the sample cross section in two-beam condition with $\mathbf{g} = 0002$ at 200\,kV. The Si substrate is located at the bottom. It can be seen that the NWs fluctuate in diameter, length and tilt. In the center of figure \ref{fig3} (b) the complete heterostrucure of one individual NW is presented. The six-fold periodic structure of the axial In$_x$Ga$_{1-x}$N/GaN MQW is visible. The single period highlighted by the white box is imaged by high-resolution TEM as depicted in figure \ref{fig3} (c).
Geometric phase analysis (GPA) reveals the In-induced strain in c-direction as shown in figure \ref{fig3} (d) \cite{Knelangen2011}. The strain $\Delta c/c$ is indicated with respect to the GaN barrier. It can be seen that it is reasonable to assume in the following the nominal value of $d_\mathrm{well} = 5$\,nm. A line profile of $\Delta c/c$ along the NW axis was obtained from the area in the GPA map highlighted by the black box, and is given in figure \ref{fig3} (e).
We would like to emphasize that due to the large ensemble fluctuations it is virtually impossible to obtain statistically relevant information about the average In content of the NW ensemble from such images. That is why our structural analysis focuses on techniques that probe the entire NW ensemble.

%%%%%%%%%%%%%%%
\subsection{X-ray simulation}
%%%%%%%%%%%%%%%

Using the method described in \cite{Kaganer2011a}, we
simulated the XRD profiles, and the resulting curves are also shown in
figure \ref{fig2}.
To account for the broadening of the satellite peaks, a standard deviation of the segment heights $\sigma_d = 2.5$\,nm has to be assumed. This rather large value reflects the inhomogeneity in heterostructure formation that is intrinsic to ensembles of self-induced nanowires grown by MBE \cite{Woelz2011b}. For the present study, it is a decisive advantage that XRD probes a large number of NWs and directly reveals the ensemble average and standard deviation. Compared to the QW thickness fluctuation within one sample of the series, the change in QW thickness between the different samples caused by different In concentrations is negligible.
The composition of the QWs $x$ is given by
$d_\mathrm{well} x = d_\mathrm{SL} x_\mathrm{avg}$.
Figure~\ref{fig4} (a) shows $x$ for $d_\mathrm{well} = 5.0$\,nm.
$x$ can be seen to decrease with increasing $T_a$.
The estimates for $x$ resulting from the limiting cases for the QW thickness, $d_\mathrm{well} \pm \sigma_d$, i.\,e. between 2.5\,nm and 7.5\,nm, are shown as error bars. 
For the sample grown at $T_a = 608$\,$^{\circ}$C the strain $\Delta c/c$, as obtained from the XRD data, is marked for comparison in the strain profile from GPA in figure \ref{fig3} (e). The solid blue line corresponds to $d_\mathrm{well} = 5.0$\,nm, $\Delta c/c=1.75$\,\% and $x=11.5$\,\%. The dashed lines give the estimates from XRD for the limiting QW thickness estimates as discussed above. Within the large experimental error, the data agree between this individual NW in HRTEM and the NW ensemble in XRD.

%%%%%%%%%%%%%%%
\subsection{Raman spectroscopy}
%%%%%%%%%%%%%%%

\begin{figure}
\hspace{1in}\includegraphics{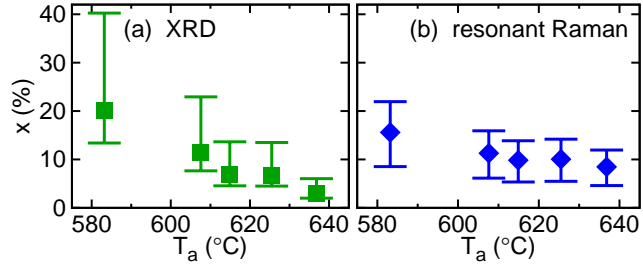}
\caption{\label{fig4}In content $x$ in In$_x$Ga$_{1-x}$N/GaN NW axial superlattices grown at different temperatures $T_a$.
(a) $x$ from XRD satellite peak positions.
(b) $x$ derived from $E_1$(LO) phonon frequency in the Raman spectra. The uncertainty ranges are described in the main text.}
\end{figure}

\begin{figure}
\hspace{1in}\includegraphics{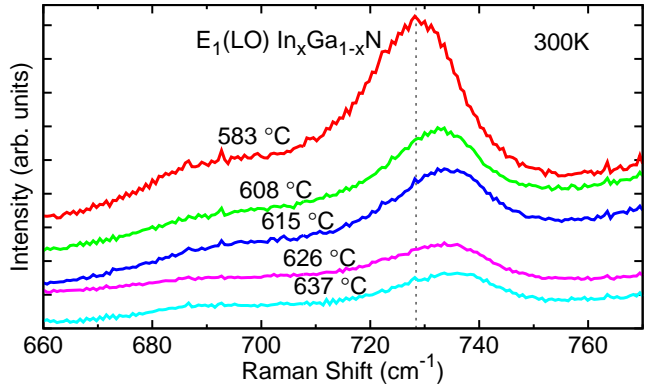}
\caption{\label{fig5}Raman spectra excited at 3.0\,eV for In$_x$Ga$_{1-x}$N/GaN nanowire axial superlattices grown at different temperatures $T_a$. The dashed line marks the $E_1$(LO) phonon frequency peak position of the sample with the highest In content.}
\end{figure}

A complementary estimation of $x$ was achieved by
resonant Raman spectroscopy \cite{Limbach2011}. The frequency of LO phonons in
In$_x$Ga$_{1-x}$N depends on both the In content and
strain \cite{Correia2003}. In order to selectively enhance the Raman
signal from the In$_x$Ga$_{1-x}$N QWs, we excited the spectra with a
photon energy of 3.0\,eV, i.\,e., close to the resonance with the
fundamental band gap for $x \approx 10$\,\%. The efficiency of scattering by LO
phonons via the Fröhlich mechanism can be considerably enhanced by
choosing this kind of resonant Raman spectroscopy \cite{Lazic2005}.
In fact, the resonantly enhanced LO-phonon signal from the In$_x$Ga$_{1-x}$N QWs is much stronger than that from the GaN base material. Using non-resonant conditions, Raman spectra from In$_x$Ga$_{1-x}$N would not be detectable because of the small scattering volume provided by the thin QWs and the small separation between In$_x$Ga$_{1-x}$N and GaN phonon frequencies at low In contents.
The corresponding spectra of our samples are depicted in figure \ref{fig5}. They are dominated by the $E_1$(LO) phonon peak. The shoulder at 685\,cm$^{-1}$ might be attributed to the $S$~band which is associated in literature with disorder or defects \cite{Correia2003,Kontos2005}.
Contrary to planar layers, light is coupled into and out of c-plane In$_x$Ga$_{1-x}$N laterally through the side facets (perp. to the c-axis) in our as-grown ensembles of vertical NWs, and scattering by $E_1$(LO) phonons dominates over that by $A_1$(LO) phonons \cite{Lazic2008a}.
Since the $E_1$(LO) frequency shift with In content and strain is not well documented, we assumed that the offset against $A_1$(LO) phonons is independent of the In content and strain, and used published dependencies for $A_1$(LO). With decreasing $T_a$, the peak shifts towards
lower frequencies. This shift amounts to $-149 x$\,cm$^{-1}$ for
relaxed In$_x$Ga$_{1-x}$N \cite{Correia2003, Grille2000}.
Biaxial compressive strain induces higher LO phonon frequencies, and we derive a shift of $-58 x$\,cm$^{-1}$ for In$_x$Ga$_{1-x}$N grown pseudomorphically on GaN from the data given in \cite{Demangeot2004} for GaN. These limiting cases are indicated
by the error bars in figure \ref{fig4} (b). The
values of $x$ plotted in this diagram were derived under the assumption that the active region relaxes laterally and the relative
strain is equal to $d_{\mathrm{barrier}}/d_{\mathrm{SL}}$ as extracted from the XRD measurements.
In agreement with the XRD results, $x$ decreases with increasing $T_a$.

The range of In content deduced from the Raman measurements is smaller than the range derived from XRD. This difference could arise from the resonant excitation which favors regions with $x \approx 10$\,\% if the composition fluctuates within a given sample. The weaker Raman signal for samples with higher
$T_a$ as seen in figure \ref{fig5} is then explained by the detuning of the resonance condition for lower In content. Another factor might be the superimposed contribution of $A_1$(LO) phonons, which would lead to an overestimation of $x$.
While for the rest of this discussion we rely on the absolute values of $x$ as derived from XRD, we do see the resonant Raman results as an independent proof of the systematic In content variation with growth temperature.

%%%%%%%%%%%%%%%
\section{Luminescence properties}
%%%%%%%%%%%%%%%
%%%%%%%%%%%%%%%
\subsection{Photoluminescence}
%%%%%%%%%%%%%%%

\begin{figure}
\hspace{1in}\includegraphics{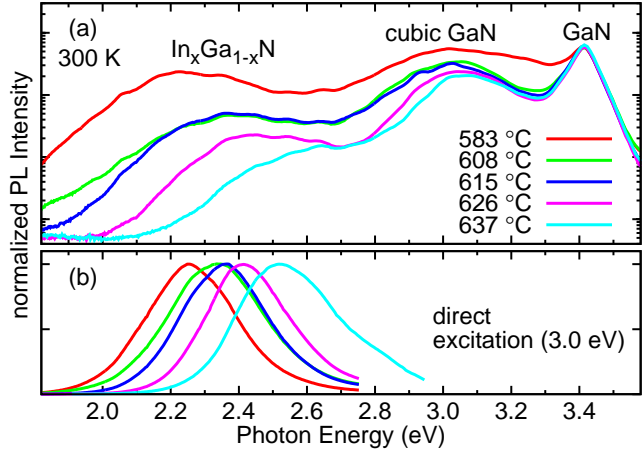}
\caption{\label{fig6}Room temperature PL spectra of In$_x$Ga$_{1-x}$N/GaN NW axial superlattices grown at different temperatures $T_a$. (a) Excitation at 3.8\,eV (logarithmic). (b) Direct excitation at 3.0\,eV (linear).}
\end{figure}

With the structural parameters elucidated, we turn to the emission properties. Room temperature photoluminescence (PL) spectra of the different NW ensembles are presented in figure \ref{fig6} (a) for excitation at 3.8\,eV, i.\,e. above the band gap of GaN. They show three main emission bands.
The peak at 3.4\,eV represents the free exciton transition in the strain-free GaN NW base and is common to all samples. The spectra are normalized to this peak.
The second peak at 3.0\,eV, which is observed for all samples grown at lower temperatures, is attributed to emission from insertions of cubic GaN (c-GaN) \cite{Renard2010}. Direct evidence for the formation of c-GaN was obtained \emph{in situ} by RHEED as described above.
The redshift with respect to the c-GaN room temperature band edge of 3.2\,eV may be explained by the quantum-confined Stark effect (QCSE) in thin layers of c-GaN in a matrix of hexagonal GaN \cite{Jacopin2011b}.
The third PL emission peak in the energy range from 2.2\,eV to 2.5\,eV shifts systematically with $T_a$. In view of the variation in In content with $T_a$, we attribute this emission to the In$_x$Ga$_{1-x}$N QWs. The shift in the peak position can be seen more clearly in the normalized spectra depicted in figure \ref{fig6} (b) which were obtained by direct excitation at 3.0\,eV. The intensity of the In$_x$Ga$_{1-x}$N emission decreases drastically with decreasing In content. This decrease in intensity with decreasing In content (and therefore presumably higher crystalline quality and reduced QCSE) is unexpected and contradictory to experience with planar (In,Ga)N structures.

%%%%%%%%%%%%%%%
\subsection{Calculation of transition energies}
%%%%%%%%%%%%%%%

\begin{figure}
\hspace{1in}\includegraphics{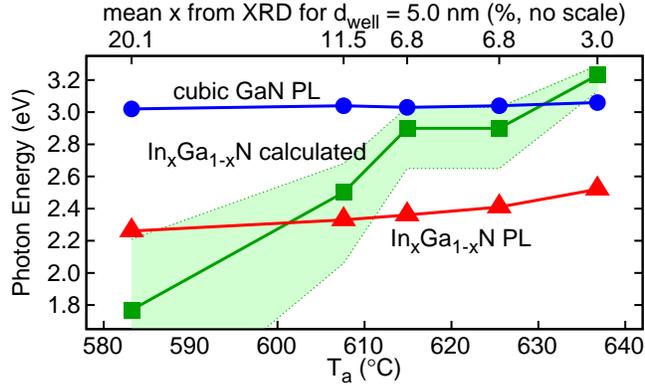}
\caption{\label{fig7}Experimental PL peak positions and calculated planar In$_x$Ga$_{1-x}$N QW transition energies. The shaded area results from the maximum and minimum estimation of $x$ and $d_\mathrm{well}$ from XRD.}
\end{figure}

Figure~\ref{fig7} shows the observed peak positions for the emission from c-GaN and the In$_x$Ga$_{1-x}$N QWs as a function of $T_a$.
To correlate the experimentally observed emission energies and structural parameters, we calculated the expected transition energies by solving the one-dimensional Poisson and Schr\"odinger equations self-consistently \cite{Snider}. As input, we used the In$_x$Ga$_{1-x}$N QW thicknesses $d_{\mathrm{well}}$ given above, compositions $x$ derived from XRD, and published values for the bowed band-gaps \cite{Schley2007}, the band-offset \cite{King2008a}, and the bowed polarization \cite{Fiorentini2002}.
The simulated transition energies are added to figure \ref{fig7} for comparison with the experimentally observed PL peak positions. The shaded range of values results from the estimates for $x$ derived from XRD with different assumptions for $d_{\mathrm{well}}$ as shown in figure \ref{fig4} (a).
The PL at 3.0 eV is of almost constant energy, which supports our assignment of this emission to c-GaN insertions. In contrast, the luminescence around 2.2--2.5\,eV, assigned to emission from In$_x$Ga$_{1-x}$N, systematically shifts to higher energies with decreasing In content. However, this shift is significantly smaller than the calculation predicts, regardless of the specific assumption for $d_\mathrm{well}$. Thus, the situation in our NW based axial In$_x$Ga$_{1-x}$N QWs differs considerably from that of planar QWs, for which one-dimensional calculations are in much better agreement with the experimental emission energies \cite{Brandt2003}. We have recently shown that the reduction in piezoelectric polarization from better strain accommodation in the NW structure is not significant enough to explain the reduced shift in emission energy \cite{Laehnemann2011a}.

%%%%%%%%%%%%%%%
\section{Discussion}
%%%%%%%%%%%%%%%

Both the luminescence energy range we observe and the absence of bright emission above 2.7\,eV normally expected for low In content are in accord with similar optical measurements in other reports where the In content was not determined independently \cite{Kikuchi2006, Guo2010, Bavencove2010, Armitage2010, Lin2010, Nguyen2011a, Chang2010}. This agreement is intriguing because it indicates a universal characteristic of axial In$_x$Ga$_{1-x}$N/GaN NW heterostructures.
In principle, low PL intensity could be caused by nonradiative point defects, but it seems highly unlikely that the density of these hypothetical point defects increases with decreasing In content at higher growth temperatures. Moreover, point defects would not affect the emission energy of the In$_x$Ga$_{1-x}$N/GaN heterostructures. As a more probable explanation, three types of carrier localization have been discussed to influence the emission energy and intensity of such heterostructures. First, exciton localization induced by compositional fluctuations was observed in similar samples \cite{Chang2010,Laehnemann2011a,Bardoux2009}. Spatially direct exciton localization alone, however, would explain neither the low absolute intensity nor the cross-over between the observed PL energy and calculated transition energy for the highest In content seen in figure \ref{fig7}. Second, we have recently seen that individual electron and hole localization in separate potential minima with varying distance plays a role for In$_x$Ga$_{1-x}$N QWs in GaN NWs \cite{Laehnemann2011a}. A spatial separation of electrons and holes implies a reduced oscillator strength, thus accounting for the low PL intensity. If this separation occurs mainly in the lateral direction which is not affected by the QCSE, the redshift could be reduced compared with the one-dimensional calculations in axial direction.
Third, a three-dimensional numerical analysis \cite{Boecklin2010} indicates that axial In$_x$Ga$_{1-x}$N/GaN NW heterostructures exhibit laterally inhomogeneous strain resulting in band bending and a lateral separation of electrons and holes. Likewise, this could explain the low PL intensity. Moreover, numerical studies for axial GaN/(Al,Ga)N QWs in NWs demonstrate that both geometry and chemical composition significantly affect the location of the potential minima in the conduction and valence band leading to carrier localization \cite{Rivera2007,Rigutti2010a}. Further, our TEM images as well as those found in the literature \cite{Armitage2010,Chang2010,Knelangen2011} show that the In$_x$Ga$_{1-x}$N QWs are laterally embedded in a GaN shell in contrast to the structure assumed in \cite{Boecklin2010}. Thus, although the detailed findings of \cite{Boecklin2010} on where electrons and holes are localized cannot be generalized, the inhomogeneous strain could possibly account for the cross-over between the results of the one-dimensional model and the observed PL.

In conclusion, our results show a clear trend of emission energy with In incorporation, even though it is weaker than expected. In contrast to planar In$_x$Ga$_{1-x}$N/GaN heterostructures, the analysis of the optical transitions in NW heterostructures necessitates---even for a rather basic level---fully three-dimensional simulations that take into account lateral variations in strain. Our findings suggest, however, that the interplay of several mechanisms for carrier localization has to be considered to describe the luminescence from NW based In$_x$Ga$_{1-x}$N/GaN heterostructures.

%%%%%%%%%%%%%%%
\ack % Acknowledgements
%%%%%%%%%%%%%%%

We are grateful to C. Herrmann, G. Jaschke, and H.-P. Schönherr for their dedicated maintenance of the MBE system and would like to thank Greg Snider for a scriptable adaptation of his 1D Poisson program. This work has been partly funded by the German government BMBF project MONALISA (Contract no. 01BL0810).

%%%%%%%%%%%%%%%
\section*{References}
%%%%%%%%%%%%%%%

\bibliographystyle{iopart-num}
\bibliography{InGaN_NW_T-Serie}

\providecommand{\newblock}{}
\begin{thebibliography}{10}
\expandafter\ifx\csname url\endcsname\relax
  \def\url#1{{\tt #1}}\fi
\expandafter\ifx\csname urlprefix\endcsname\relax\def\urlprefix{URL }\fi
\providecommand{\eprint}[2][]{\url{#2}}

\bibitem{Kikuchi2006}
Kikuchi A, Tada M, Miwa K and Kishino K 2006 {\em Proc. SPIE\/} {\bf 6129}
  612905

\bibitem{Guo2010}
Guo W, Zhang M, Banerjee A and Bhattacharya P 2010 {\em Nano Lett.\/} {\bf 10}
  3355

\bibitem{Bavencove2010}
Bavencove A~L, Tourbot G, Pougeoise E, Garcia J, Gilet P, Levy F, André B,
  Feuillet G, Gayral B, Daudin B and Dang L~S 2010 {\em Phys. Stat. Sol. A\/}
  {\bf 207} 1425

\bibitem{Armitage2010}
Armitage R and Tsubaki K 2010 {\em Nanotechnology\/} {\bf 21} 195202

\bibitem{Lin2010}
Lin H~W, Lu Y~J, Chen H~Y, Lee H~M and Gwo S 2010 {\em Appl. Phys. Lett.\/}
  {\bf 97} 073101

\bibitem{Nguyen2011a}
Nguyen H~P~T, Cui K, Zhang S, Fathololoumi S and Mi Z 2011 {\em
  Nanotechnology\/} {\bf 22} 445202

\bibitem{Geelhaar2011}
{Geelhaar L \it{et al}} 2011 {\em IEEE J. Sel. Top. Quantum Electron.\/} {\bf
  17} 878

\bibitem{Bjoerk2002}
Björk M~T, Ohlsson B~J, Sass T, Persson A~I, Thelander C, Magnusson M~H,
  Deppert K, Wallenberg L~R and Samuelson L 2002 {\em Appl. Phys. Lett.\/} {\bf
  80} 1058

\bibitem{Chang2010}
Chang Y~L, Wang J~L, Li F and Mi Z 2010 {\em Appl. Phys. Lett.\/} {\bf 96}
  013106

\bibitem{Knelangen2011}
Knelangen M, Hanke M, Luna E, Schrottke L, Brandt O and Trampert A 2011 {\em
  Nanotechnology\/} {\bf 22} 365703

\bibitem{Kaganer2011a}
Kaganer V~M, Wölz M, Brandt O, Geelhaar L and Riechert H 2011 {\em Phys. Rev.
  B\/} {\bf 83} 245321

\bibitem{Woelz2011b}
Wölz M, Kaganer V~M, Brandt O, Geelhaar L and Riechert H 2011 {\em Appl. Phys.
  Lett.\/} {\bf 98} 261907

\bibitem{Woelz2012}
Wölz M, Kaganer V~M, Brandt O, Geelhaar L and Riechert H 2012 {\em Appl. Phys.
  Lett.\/} {\bf 100} 179902

\bibitem{Averbeck1999}
Averbeck R and Riechert H 1999 {\em Phys. Stat. Sol. A\/} {\bf 176} 301

\bibitem{Kaganer2012}
Kaganer V~M and Belov A~{\mbox{Yu}} 2012 {\em Phys. Rev. B\/} {\bf 85} 125402

\bibitem{Limbach2011}
Limbach F, Gotschke T, Stoica T, Calarco R, Sutter E, Ciston J, Cuscó R, Artus
  L, Kremling S, Höfling S, Worschech L and Grützmacher D 2011 {\em J. Appl.
  Phys.\/} {\bf 109} 014309

\bibitem{Correia2003}
Correia M~R, Pereira S, Pereira E, Frandon J and Alves E 2003 {\em Appl. Phys.
  Lett.\/} {\bf 83} 4761

\bibitem{Lazic2005}
Lazić S, Moreno M, Calleja J~M, Trampert A, Ploog K~H, Naranjo F~B, Fernandez
  S and Calleja E 2005 {\em Appl. Phys. Lett.\/} {\bf 86} 061905

\bibitem{Kontos2005}
Kontos A~G, Raptis Y~S, Pelekanos N~T, Georgakilas A, Bellet-Amalric E and
  Jalabert D 2005 {\em Phys. Rev. B\/} {\bf 72} 155336

\bibitem{Lazic2008a}
Lazić S, Gallardo E, Calleja J~M, Agullo-Rueda F, Grandal J, Sanchez-Garcia
  M~A and Calleja E 2008 {\em Physica E\/} {\bf 40} 2087

\bibitem{Grille2000}
Grille H, Schnittler C and Bechstedt F 2000 {\em Phys. Rev. B\/} {\bf 61} 6091

\bibitem{Demangeot2004}
Demangeot F, Frandon J, Baules P, Natali F, Semond F and Massies J 2004 {\em
  Phys. Rev. B\/} {\bf 69} 155215

\bibitem{Renard2010}
Renard J, Tourbot G, Sam-Giao D, Bougerol C, Daudin B and Gayral B 2010 {\em
  Appl. Phys. Lett.\/} {\bf 97} 081910

\bibitem{Jacopin2011b}
Jacopin G, Rigutti L, Largeau L, Fortuna F, Furtmayr F, Julien F~H, Eickhoff M
  and Tchernycheva M 2011 {\em J. Appl. Phys.\/} {\bf 110} 064313

\bibitem{Snider}
Snider G {1D Poisson} freeware University of Notre Dame
  \urlprefix\url{http://www.nd.edu/\~gsnider/}

\bibitem{Schley2007}
Schley P, Goldhahn R, Winzer A~T, Gobsch G, Cimalla V, Ambacher O, Lu H, Schaff
  W~J, Kurouchi M, Nanishi Y, Rakel M, Cobet C and Esser N 2007 {\em Phys. Rev.
  B\/} {\bf 75} 205204

\bibitem{King2008a}
King P~D~C, Veal T~D, Kendrick C~E, Bailey L~R, Durbin S~M and McConville C~F
  2008 {\em Phys. Rev. B\/} {\bf 78} 033308

\bibitem{Fiorentini2002}
Fiorentini V, Bernardini F and Ambacher O 2002 {\em Appl. Phys. Lett.\/} {\bf
  80} 1204

\bibitem{Brandt2003}
Brandt O, Sun Y~J, Schönherr H~P, Ploog K~H, Waltereit P, Lim S~H and Speck
  J~S 2003 {\em Appl. Phys. Lett.\/} {\bf 83} 90

\bibitem{Laehnemann2011a}
Lähnemann J, Brandt O, Pfüller C, Flissikowski T, Jahn U, Luna E, Hanke M,
  Knelangen M, Trampert A and Grahn H~T 2011 {\em Phys. Rev. B\/} {\bf 84}
  155303

\bibitem{Bardoux2009}
Bardoux R, Kaneta A, Funato M, Kawakami Y, Kikuchi A and Kishino K 2009 {\em
  Phys. Rev. B\/} {\bf 79} 155307

\bibitem{Boecklin2010}
Böcklin C, Veprek R~G, Steiger S and Witzigmann B 2010 {\em Phys. Rev. B\/}
  {\bf 81} 155306

\bibitem{Rivera2007}
Rivera C, Jahn U, Flissikowski T, Pau J~L, Munoz E and Grahn H~T 2007 {\em
  Phys. Rev. B\/} {\bf 75} 045316

\bibitem{Rigutti2010a}
Rigutti L, Teubert J, Jacopin G, Fortuna F, Tchernycheva M, Bugallo A~D, Julien
  F~H, Furtmayr F, Stutzmann M and Eickhoff M 2010 {\em Phys. Rev. B\/} {\bf
  82} 235308

\end{thebibliography}

\end{document}